\documentclass[aps,prd,superscriptaddress,nofootinbib,amsfonts,amssymb,amsmath,notitlepage,twocolumn]{revtex4-1}

\usepackage[dvipdfmx]{graphicx}
\usepackage{amsmath}
\usepackage{amssymb} \allowdisplaybreaks[0]
\usepackage{float}
\usepackage[colorlinks=true, allcolors=blue!75!black]{hyperref}
\usepackage{makecell}
\usepackage{multirow}
\usepackage{slashed}
\usepackage{wrapfig}
\usepackage{bm}
\usepackage{color}
\usepackage{xcolor}
\usepackage[nameinlink]{cleveref}

\newcommand{\Slash}[1]{{\ooalign{\hfil/\hfil\crcr$#1$}}} 
\newcommand{\bvec}[1]{\mbox{\boldmath $#1$}}

\newcommand{\df}{\text{d}}

\graphicspath{{./figs/}}
\newbox{\ORCIDicon}
\sbox{\ORCIDicon}{\large
                  \includegraphics[width=0.8em]{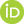}}

\begin{document}

\title{
The effect of charm quark on the QCD chiral phase diagram  
}

\author{Fei \surname{Gao}\,\href{https://orcid.org/0000-0001-5925-5110}{\usebox{\ORCIDicon}}}
\affiliation{School of Physics, Beijing Institute of Technology, 100081 Beijing, China}

\author{Yuepeng \surname{Guan}\,\href{https://orcid.org/0009-0007-8571-0931}{\usebox{\ORCIDicon}}}
\affiliation{Center for Theoretical Physics and College of Physics, Jilin University, Changchun, 130012, China}

\author{Shinya \surname{Matsuzaki}\,\href{https://orcid.org/0000-0003-4531-0363}{\usebox{\ORCIDicon}}}
\affiliation{Center for Theoretical Physics and College of Physics, Jilin University, Changchun, 130012, China}

\begin{abstract}
We study the influence of charm-quark dynamics on the chiral phase structure of Quantum Chromodynamics (QCD) using the recently developed miniDSE scheme for the Dyson–Schwinger equations. 
By comparing the quark and gluon propagators in the $2+1$- and $2+1+1$-flavor setups within the same truncation scheme, we qualify the impact of the charm-quark loop on the QCD phase diagram. 
Our results show that the charm quark has only a mild effect on the crossover boundary, which remains almost unchanged within the present setup. 
The most visible effect is a small shift of the critical endpoint (CEP) toward lower baryon chemical potential, by approximately $3\%$.
The present result provides a controlled estimate of the charm-loop effect within the same miniDSE truncation. 
It indicates that heavy-flavor contributions may become relevant when aiming at precision studies of the CEP location.
\end{abstract}
\maketitle

\section{Introduction}

Quantum Chromodynamics (QCD), the fundamental theory governing the strong interaction, exhibits a rich and intricate phase structure. 
The most intriguing feature there is the spontaneous breaking and the restoration of chiral symmetry, which underlies the dynamical generation of baryon masses.
A detailed understanding of the chiral phase transition is not only essential for deciphering the nonperturbative dynamics of QCD but also has profound implications for a wide range of physical phenomena, including the physics of heavy-ion collisions~\cite{STAR:2010vob,Mohanty:2011nm,Luo:2017faz,Nonaka:2019fad,Chen:2024aom}, the equation of state of neutron stars~\cite{Buballa:2003qv,Fukushima:2013rx,Oertel:2016bki}, and the thermal evolution of the early universe~\cite{Planck:2018vyg,Gao:2021nwz,Zheng:2024tib,Gao:2024fhm}.

Over the past decades, considerable efforts have been devoted to exploring the QCD chiral phase structure, particularly in systems with two or two-plus-one quark flavors.
These investigations span a variety of nonperturbative techniques, including effective models and extrapolation methods~\cite{Shao:2011fk,Xin:2014ela,He:2013qq,Chelabi:2015gpc,Kojo:2020ztt,Chen:2020ath,Hippert:2023bel,Cai:2022omk,Chen:2018vty,Basar:2023nkp,Adam:2025pii,Ecker:2025vnb}, and first principle QCD approaches like lattice QCD simulation~\cite{Borsanyi:2020fev,HotQCD:2018pds,Bonati:2018nut}, functional QCD (fQCD) methods including the Dyson-Schwinger equations (DSEs)~\cite{Qin:2010nq,Fischer:2014ata,Gao:2016qkh,Fischer:2018sdj,Gao:2020qsj,Gao:2020fbl,Gunkel:2021oya} and functional Renormalization Group (fRG) approach~\cite{Fu:2019hdw,Dupuis:2020fhh,Fu:2022gou}.  
These studies have converged on a picture in which, at vanishing baryon density, the chiral transition for light quarks is a smooth crossover occurring at a pseudo-critical temperature of approximately $T_c \sim 150 \, \text{MeV}$.
At finite density, evidence points toward the existence of a critical endpoint (CEP) at a baryon chemical potential around $\mu_B^{\text{CEP}} \sim 600 \, \text{MeV}$. The prediction for the CEP location is particularly important as it may shed light on the precision measurement in the heavy ion collision experiments~\cite{Chen:2024aom,STAR:2025zdq}.

Most of the theoretical calculations on QCD phase transitions, however, neglect the contribution of heavy quark flavors, particularly charm and bottom, based on the decoupling theorem, which asserts that heavy flavors should have little influence on the infrared (IR) dynamics of QCD.  Nevertheless, the heavy quark dynamics including the charm fluctuations, the heavy quark drag and diffusion coefficients are important for the related phenomena in heavy ion collisions~\cite{vanHees:2005wb, Goswami:2024hfg}. 
In the pioneering works within a simplified truncation scheme of DSEs, it is found that the CEP location is barely changed with the inclusion of the charm quark loop~\cite{Fischer:2014ata, Welzbacher:2014pea}. 

The present work revisits the influence of charm quark dynamics on the QCD chiral phase structure using the recently proposed miniDSE approach~\cite{Lu:2023mkn}. 
This framework offers a numerically efficient and systematically improvable formulation of the Dyson–Schwinger equations, allowing for the seamless inclusion of additional quark flavors with controlled approximations and quantified uncertainties.
Within this setting, we compute and compare the chiral phase diagrams of QCD with and without the charm-quark loop in the same miniDSE truncation. 
This direct comparison shows that the crossover boundary is almost unchanged, while the CEP is shifted mildly toward lower baryon chemical potential. 
We further identify this shift with the charm-induced suppression of the gluon dressing function at intermediate momenta. 
The present work, therefore, provides an estimate of the size of the charm-loop effect on the QCD phase structure within a fixed functional framework.

This paper is arranged as follows.
In Sec.~\ref{sec:QCDSetups}, we briefly introduce the quark and gluon propagator DSEs of QCD and summarize the truncation scheme in the current work.
In Sec.~\ref{sec:Results}, we show the impacts of the charm quark loop on the physical observables in vacuum.
Then, we present the chiral phase diagram obtained from the different cases, and list the observed quantities from the phase diagrams.
The conclusions and discussions are put in Sec.~\ref{sec:Conclusion}.

\section{Dyson-Schwinger equations at finite temperature and chemical potential}
\label{sec:QCDSetups}

In the current work, we employ the computationally minimal truncation scheme for the DSE approach (miniDSE) of QCD provided in Ref.~\cite{Lu:2023mkn}.
The strategy is to solve the quark gap equations self-consistently, and deal with the difference of DSE for the gluon propagator with respect to the input data coming from other quantitative approaches.
The coupled equation would be closed after we impose the dressing functions for the quark-gluon vertices with leading-order tensorial structures, i.e., the Dirac and Pauli terms.

We formulate the DSEs in the Landau gauge and parameterize the quark and gluon propagators as
\begin{gather}
    \big[ S^f(\tilde p) \big]^{-1} = i \tilde \omega_{n}^\psi \gamma_4 C^f(\tilde p) + i \bvec{p} \cdot \bvec{\gamma} A^f(\tilde p) + B^f(\tilde p), \nonumber\\
    D_{\mu\nu}(k) = G_A(k) \Pi^\perp_{\mu\nu},
    \label{eq:parameterizationOfPropagators}
\end{gather}
where the flavor index takes $f = u,d,s,c$.
In Eq.~\eqref{eq:parameterizationOfPropagators}, the scalar part of the gluon propagator reads
\begin{align}
    G_A(k) = \frac{Z_A(k)}{k^2}.
\end{align}
We adopt the $O(4)$-symmetric approximation for the gluon propagator such that the chromoelectric and the chromomagnetic parts do not split from each other.
The momentum arguments of quarks and gluon are given by $\tilde p_\mu = (\tilde \omega_{n}^\psi, \bvec{p})$ and $k = (\omega_{n}^A, \bvec{k})$, with $\tilde \omega_{n}^\psi = \omega_{n}^\psi + i \mu_q = (2 n + 1) \pi T$ and $\omega_{n}^A = 2 n \pi T$.
The quark chemical potential takes $\mu_q = \mu_B/3$ with the flavor-universal baryon chemical potential $\mu_B$.

\subsection{The quark gap equation}

The quark gap equations for each flavor read
\begin{align}
    \big[ S^f(\tilde p) \big]^{-1} &= \big[ S_0^f(\tilde p) \big]^{-1} + \Sigma_0^f(\tilde p),
    \label{eq:quarkGapEquation}
\end{align}
where the expression of quark self-energy $\Sigma_0^f$ is given by
\begin{align}
    \Sigma_0^f(\tilde p) = \frac{4}{3} Z_1^f g_s \int_q \, \gamma_\mu S^f(\tilde q) \Gamma^f_{\nu}(\tilde q, \tilde p) D_{\mu\nu}(k),
\end{align}
where $k_\mu = \tilde q - \tilde p$, with the shorthand notation $\int_q = T\sum_n \int \df^3 \bvec{k}$.
Following the miniDSE construction~\cite{Lu:2023mkn}, we pick up the first and the fourth transverse tensor structure such that
\begin{align}
    \Gamma^f_{\mu}(\tilde q, \tilde p) = \mathcal T_\mu^{f,(1)}(\tilde q, \tilde p) \lambda^{(1)}(\tilde q, \tilde p) + \mathcal T_\mu^{(4)}(\tilde q, \tilde p) \lambda^{f,(4)}(\tilde q, \tilde p),
\end{align}
with
\begin{gather}
    \mathcal T_\mu^{(1)}(\tilde q, \tilde p) = -i \gamma_\mu, \quad \mathcal T_\mu^{(4)}(\tilde q, \tilde p) = i \sigma_{\mu\nu} k_\nu, \nonumber\\
    \sigma_{\mu\nu} = \frac{i}{2} \big[ \gamma_\mu, \gamma_\nu \big].
\end{gather}
In the vertex structure, the Dirac structure $\mathcal T_\mu^{(1)}$ is the tree-level structure and is naturally dominant. The Pauli term $\mathcal T_\mu^{(4)}$ is important in the perturbative calculation, and is found to be also dominant in the nonperturbative calculations~\cite{Tang:2019zbk,Gao:2021wun}. Now with the parameterization given in Eq.~\eqref{eq:parameterizationOfPropagators}, the dressing function of the Dirac term is given by
\begin{align}
    \lambda^{f,(1)}(\tilde q, \tilde p) = g_s F(k^2) \Big[ \delta_{\mu 4} \Sigma_{C^f}(\tilde q, \tilde p) + (1-\delta_{\mu 4}) \Sigma_{A^f}(\tilde q, \tilde p) \Big]
    \label{eq:lambda1forvertex}
\end{align}
to satisfy the Slavnov-Taylor identity (STI) for the quark-gluon vertex, where $F(k^2) = k^2 G_c(k)$ denotes the ghost dressing function, and $G_c(k) \delta^{ab}$ is the ghost propagator.

Also, we employ
\begin{align}
    \lambda^{f,(4)}(\tilde q, \tilde p) = g_s Z_A^{-\frac{1}{2}}(k) \Delta_{B^f}(\tilde q, \tilde p)
\end{align}
following the construction in Ref.~\cite{Lu:2023mkn}.
The functions $\Sigma_{X}$ and $\Delta_X$ are defined as
\begin{gather}
    \Sigma_{X}(\tilde q, \tilde p) \equiv \frac{X(\tilde q) + X(\tilde p)}{2}, \nonumber\\
    \Delta_{X}(\tilde q, \tilde p) \equiv \frac{X(\tilde p) - X(\tilde q)}{\tilde p^2 - \tilde q^2},
\end{gather}
with $X(\tilde p)$ is an arbitrary function of the fermionic momentum argument $\tilde p$.

The information of the ghost dressing function $F(k^2)$ is also needed to evaluate the vertex dressing function \eqref{eq:lambda1forvertex}.
In the current work, we take the quantitative result from another functional approach as the input to the ghost dressing function at vacuum, and drop the temperature and baryon chemical potential dependence as an approximation, which is shown to be a milder influence on the current result.
The fit function of the ghost dressing function is given by
%
\begin{align}
    F(k^2) = \frac{a_1 + b_1 \sqrt{k^2} + c_1 k^2}{1 + d_1 \sqrt{k^2} + e_1 k^2},
\end{align}
where the values of the parameters are chosen as~\cite{Aguilar:2021okw}
%
\begin{align}
&a_1 = 3.169~{\rm GeV^2},&
&b_1 = 4.744~{\rm GeV},&
&c_1 = 6.045,\nonumber\\
&d_1 = 1.558~{\rm GeV^2},&
&e_1 = 6.385~{\rm GeV}.
\end{align}
This then closes the quark gap equation, and the quark propagator can be obtained after putting in the gluon propagator.

\subsection{The gluon gap equation}


For the gluon gap equation, we take the following difference form of DSE~\cite{Gao:2020qsj, Lu:2023mkn}
\begin{align}
    D_{\mu\nu,\bvec{v}}^{-1}(k) = D_{\mu\nu,\bvec{v}_0}^{-1}(k) + \Delta \Pi^{\bvec{v},\bvec{v}_0}_{A,\mu\nu}(k),
    \label{eq:diffDSEforGluonProp}
\end{align}
and here the difference is for the inclusion of the charm quark, which then reads:
\begin{gather}
    \bvec{v}_0 = \Big( N_f = 2+1, m_f = (m_l,m_l,m_s), T, \mu_q \Big), \nonumber\\
    \bvec{v} = \Big( N_f = 2+1+1, m_f = (m_l,m_l,m_s,m_c), T, \mu_q \Big).
\end{gather}
For the gluon propagator with $2+1$ flavor structure $D_{\mu\nu,\bvec{v}_0}(k) = Z_{A,\bvec{v}_0}(k)/k^2 \cdot \Pi^\perp_{\mu\nu}$, we use the hard thermal loop approximation to approach the quantitative result, which reads
\begin{align}
    Z_{A,\bvec{v}_0}(k) = Z^{\rm trans}_{T,\mu_q}(k),
\end{align}
where the right-hand side denotes the transverse gluon propagator dressing at finite temperature and quark chemical potential, and 
\begin{align}
    Z^{{\rm trans}}_{T,\mu_q}(k) = Z_{\rm vac}(p) + \frac{2\pi}{3} \alpha_S^\text{HTL}  \Biggl( T^2 + \frac{\mu_B^2}{3 \pi^2} \Biggl) \frac{1}{k^2},
    \label{eq:HTLmassFromQuarkLoop}
\end{align}
%

\begin{figure}[t]
    \centering
    \includegraphics[width=1\linewidth]{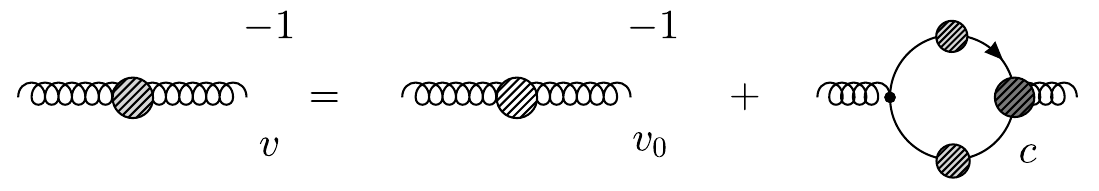}
    \caption{
    Feynman diagrams of the difference DSE~\eqref{eq:diffDSEforGluonProp}.
    The black-solid lines with gray-hatched blobs represent the charm quark propagators, and the spiraling lines with gray- or white-hatched blobs denote the gluon propagator in $2+1$ or $2+1+1$ flavors. 
    }
    \label{fig:DiffDSEforGLuon}
\end{figure}

where $\alpha_S^\text{HTL} = 0.115$ is the gauge coupling for evaluating the hard thermal loop (HTL) mass.
In Eq.~\eqref{eq:HTLmassFromQuarkLoop}, we only take into account the light quark loop contribution to the gluon thermal mass, and the strange flavor contribution is dropped due to its heaviness. 
For the vacuum gluon dressing $Z^{-1}_{\rm vac}(p)$, we take the following propagator form \cite{Gao:2021wun, Lu:2023mkn}
\begin{align}
    Z_{\rm vac}^{-1}(k^2) = \frac{k^2\frac{a^2 + k^2}{b^2 + k^2}}{M_G^2(k^2) + k^2 \left[ 1 + c \log (d^2 k^2 + e^2 M_G^2(k^2)) \right]^\gamma},
    \label{eq:gluonPropagator}
\end{align}
in which
\begin{align}
    M_G^2(k^2) = \frac{f^4}{g^2 + k^2},
    \label{eq:gluonMass}
\end{align}
and $\gamma = \frac{13-\frac{4}{3} N_f}{22 - \frac{4}{3}N_f}$ is the perturbative anomalous dimension of the gluon propagator.
Here we choose $N_f = 3$ to match the $(2+1)$-flavor result of the gluon propagator.
The values of the parameters are chosen as 
\begin{align}
&a= 1~{\rm GeV},&
&b=0.735~{\rm GeV},&
&c= 0.12,\nonumber\\
&d=0.0257~{\rm GeV}^{-1},&
&e=0.081~{\rm GeV}^{-1}.
\end{align}
in the gluon propagator~\labelcref{eq:gluonPropagator} and 
\begin{align}
&f=0.65~{\rm GeV},&
&g=0.87~{\rm GeV},
\end{align}
in the gluon mass~\labelcref{eq:gluonMass}, to fit with the $(2+1)$-flavor data from the functional approaches~\cite{Gao:2021wun}.

The self-energy in the difference DSE~\eqref{eq:diffDSEforGluonProp} for the gluon propagator is then given by the charm quark vacuum polarization 
\begin{align}
    &\Delta \Pi^{\bvec{v},\bvec{v}_0}_{A,\mu\nu}(k) \nonumber\\
    &= -\frac{Z_1^f}{2} g_s^{\rm HTL} \int_q \, \operatorname{tr} \Big[ \gamma_\mu S^c(\tilde q) \Gamma^c_{\nu}(\tilde q, \tilde q - k) S^c(\tilde q - k) \Big].
    \label{eq:charmQuarkLoop}
\end{align}
which is diagrammatically shown in Fig.~\ref{fig:DiffDSEforGLuon}, and $g_s^{\rm HTL} = \sqrt{4\pi \alpha_s^{\rm HTL}}$.
The tensor structure $\mathcal T^{(4)}_\nu$  has been found to be negligible in the gluon self energy~\cite{Lu:2023mkn,Lu:2025cls}, and hence here for simplicity, we only keep the Dirac term in the evaluation of Eq.~\eqref{eq:charmQuarkLoop}.

\begin{figure}
    \centering
    \includegraphics[width=1\linewidth]{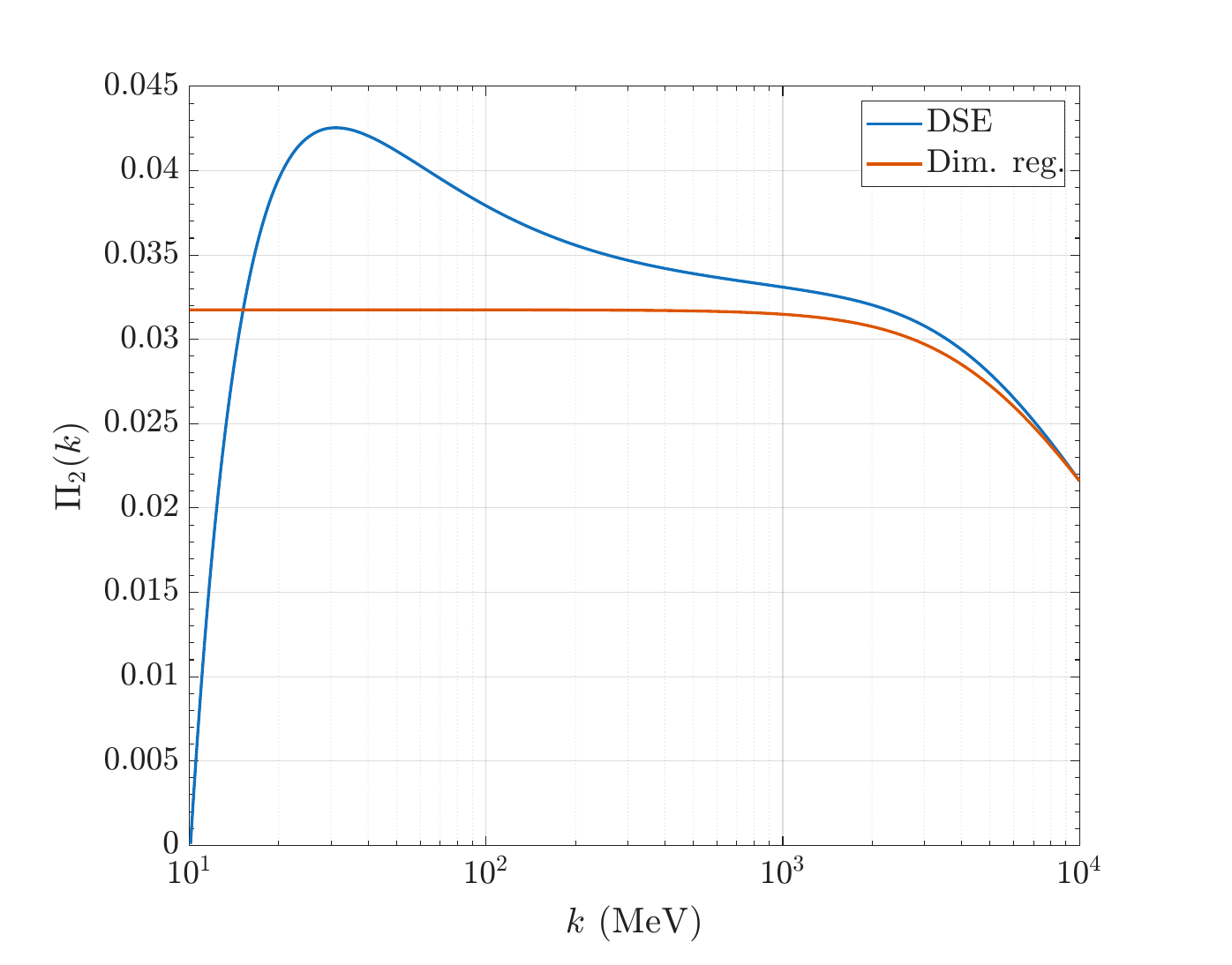}
    \caption{
    Dimensionless gluon self-energies $\Pi_2(k)$ evaluated from the DSE~\eqref{eq:charmQuarkLoop} (blue-solid line) and the dim.~reg.~\eqref{eq:AnalyticalCharmLoop} (Orange-solid line) at the vacuum.
    }
    \label{fig:CharmLoopCompare}
\end{figure}

Note that there exists quadratic divergence in Eq.~\eqref{eq:charmQuarkLoop}, which in the perturbative calculation requires the dimensional regularization. 
For the nonperturbative calculation, we perform the Brown-Pennington projection procedure to separate the divergence induced by the ultraviolet (UV) cutoff $\Lambda$ from the temperature $T$ in the quark loop integral~\cite{Brown:1988bm, Fischer:2012vc}.  The projection for the self-energy is defined as:
\begin{align}
    \Pi^{\rm B-P}_{\mu\nu} = \delta_{\mu\nu} - 4 \frac{k_\mu k_\nu}{k^2}.
\end{align}
The dimensionless gluon self-energy $\Pi_2(k)$ is obtained from the projection as
\begin{align}
    \Pi_2(k) = \frac{\Delta \Pi^{\bvec{v},\bvec{v}_0}_{A,\mu\nu}(k) \ \Pi^{\rm B-P}_{\mu\nu}(k)}{3 k^2}.
\end{align}

The quadratic divergence in the self-energy can be subtracted after this projection, and only the logarithmic divergence is left, which is absorbed by the renormalization constant. 
In comparison, we also present the benchmark result from the vacuum charm quark loop with the bare vertex and the tree-level-like propagator based on the dimensional regularization (dim.~reg.), that

\begin{align}
    \Delta \Pi^{\bvec{v},\bvec{v}_0}_{A,\mu\nu}(k) = \Pi^\perp_{\mu\nu} \cdot \Pi_{2,\rm vac}(k) \ k^2 ,
\end{align}
where the scalar part of the gluon self-energy reads~\cite{Schwartz:2014sze}
\begin{align}
    &\Pi_{2,\rm vac}(k,\mu) \nonumber\\
    &= \frac{\alpha_s^{}(\mu)}{\pi} \int_0^1 \df x \, x(1-x)\log\Big[ \frac{(m_c^2 + \mu^2 x (1-x))}{(m_c^2 + k^2 x (1-x))} \Big]  ,
    \label{eq:AnalyticalCharmLoop}
\end{align}
where $m_c$ is the current quark mass of the charm quark.
In Fig.~\ref{fig:CharmLoopCompare}, we show the dimensionless gluon self-energies $\Pi_2(k)$ from Eq.~\eqref{eq:charmQuarkLoop} by solving the coupled DSEs at the vacuum, and the one from Eq.~\eqref{eq:AnalyticalCharmLoop}.
Both agree with each other in the UV region down to the momentum scale around $k \sim 1 \, {\rm GeV}$. In the IR region, the full quark propagator and quark gluon vertex lead to an enhancement in the self-energy of the gluon propagator.



\section{Results}
\label{sec:Results}

The impact of the charm quark on the QCD phase transition, is mainly determined by the quark gap equation. Therefore, the charm quark effect is validated through the quark anti-quark self-energy loop in the gluon propagator as in Eq.~\ref{eq:diffDSEforGluonProp}. The setup in the previous section is thus complete for studying the charm quark effect, and in this section we present the numerical results.



\subsection{Charm effect in the vacuum}

The above-described scheme for DSEs has enabled one to obtain the quark and gluon propagator. The parameters required to solve the equations are only the coupling constant and the current quark masses, which can be determined by the physical quantities. After determining these parameters, we then compare the results for 2+1 and 2+1+1 flavors in vacuum.


%
\begin{figure}
    \centering
    \includegraphics[width=1\linewidth]{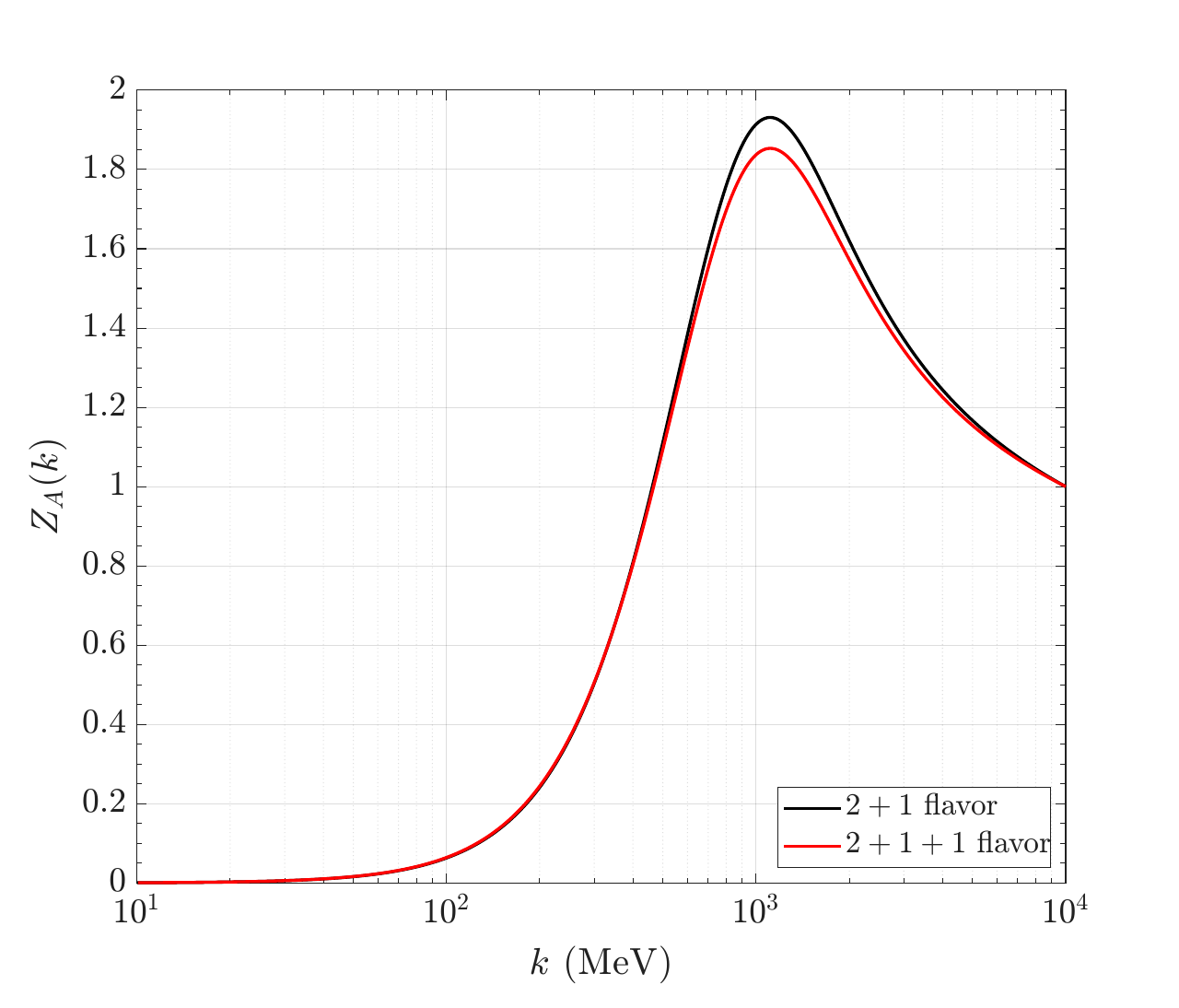}
    \caption{
    The dressing function $Z_A(k)$ of the gluon propagator in the $2+1$ flavor case (black-solid line), and in the $2+1+1$ flavor case with charm loop evaluated from DSE (red-solid line).
    }
    \label{fig:CharmMassFuncAndGluonDressing}
\end{figure}

At vanishing temperature, the quark inverse-propagator can be simply parameterized in O(4) symmetric form as
\begin{align}
    \left[ S^f(p^2) \right]^{-1} = i \Slash{p} A^f(p^2) + B^f(p^2),
\end{align}
from which we can read off the quark mass functions as $M_f(p^2) = B^f(p^2)/A^f(p^2)$, and the quark wave-function is $Z_f(p^2) = A^f(p^2)$. The quark condensate can be calculated by
\begin{align}
    \langle \bar{\psi}\psi \rangle = {4 N_c} \int_p \frac{m_l M_l^r}{Z_l(p^2 + M_l^2)},
\end{align}
with $M_l^r(p^2) = M(p^2) - m_l \partial_{m_l}M(p^2)$. 
For the $2+1$ flavor case, we take the coupling constant as $\alpha_s = 0.257$, and get
\begin{gather}
    \langle \bar{\psi}\psi \rangle ^{\frac{1}{3}} = 372.7 \, {\rm MeV}^{},\quad m_l = 1.8 \, {\rm MeV}, \quad \frac{m_s}{m_l} = 27 ,
\end{gather}
at the renormalization scale $\mu = 10 \, {\rm GeV}$.
The current quark mass of light quarks is chosen to satisfy the Gell-Mann–Oakes–Renner (GMOR) relation with the empirical values of the pion mass $m_\pi=135$ MeV
%
where the pion decay constant $f_\pi$ is evaluated through the Pagels-Stokar (PS) formula~\cite{Gao:2020qsj}
\begin{align}
    f_\pi = \frac{4N_c}{N_\pi} \int_p \frac{M_l^r}{Z_l(p^2 + M_l^2)^2} \left[ M_l - \frac{p^2}{2} \frac{\partial M_l}{\partial p^2} \right],
\end{align}
with the normalization factor $N_\pi$ of the Bethe-Salpeter wave function of the pion obtained by solving
\begin{align}
    N_\pi^2 &= f_\pi N_\pi \nonumber\\
    &\quad + 2 N_c \int_p \frac{\left( M_l^r \right)^2 \left( p^2 Z_l Z_l^{\prime \prime} + 2 Z_l Z_l^\prime - p^2Z_l^{\prime \prime} \right)}{Z_l^2(p^2 + M_l^2)},
\end{align}
where $Z_l^\prime(p^2) = \partial_{p^2}Z_l(p^2)$, $Z_l^{\prime\prime}(p^2) = \partial_{p^2}^2 Z_l(p^2)$.
With the PS formula, we get $f_\pi=101.1$ MeV, which is consistent with the previous calculation within the PS formual~\cite{Gao:2021wun}.

For the $2+1+1$ flavor case, we  put the following parameters as:
\begin{gather}
    \alpha_s = 0.255, \quad m_l = 1.8 \, {\rm MeV}, \quad \frac{m_s}{m_l} = 27,\nonumber\\
    m_c = 1270 \, {\rm MeV},
\end{gather}

The vacuum solutions, therefore, allow us to quantify the charm-flavor effect on the light-quark mass functions and on the gluon dressing function. 
After calibrating the parameters to the pion mass and decay constant, we find that the light-quark mass function is essentially unchanged by the inclusion of the charm-quark loop. 
Quantitatively, the difference between the $2+1$- and $2+1+1$-flavor results for $M_l(p)$ remains below the numerical resolution of the present calculation over the momentum range considered. 
We therefore do not display the two almost overlapping curves. 
The charm flavor has a more visible impact on the gluon dressing function $Z_A(k)$, as shown in Fig.~\ref{fig:CharmMassFuncAndGluonDressing}. 
The additional charm-quark loop mildly suppresses the gluon propagator, reducing the peak value of $Z_A(k)$ from $1.93$ in the $2+1$-flavor case to $1.85$ in the $2+1+1$-flavor case. 
Nevertheless, the characteristic momentum scale of the gluon propagator, identified with the position of the maximum of $Z_A(k)$, is barely affected. 
Toward the ultraviolet regime, the charm-induced modification gradually weakens, and the two dressing functions approach each other around $k\sim 10~\mathrm{GeV}$.

\subsection{QCD chiral phase diagram at $N_f=2+1+1$}

\begin{table}[htb]
\begin{tabular}{cccc}
\hline \hline
  & \begin{tabular}[c]{@{}c@{}}$T_c(\mu_B = 0)$\\ (MeV)\end{tabular} &  \begin{tabular}[c]{@{}c@{}}$(T^{\rm CEP}, \mu_B^{\rm CEP})$\\ (MeV)\end{tabular} \\
\hline
miniDSE~\cite{Lu:2023mkn} & $156.5$   & $(108.5, 567)$ \\
fRG~\cite{Fu:2019hdw} & $156$   & $(107, 635)$ \\
fRG-DSE: STI construction~\cite{Gao:2020qsj} & $154$   & $(93, 672)$ \\
fRG-DSE: self-consistent~\cite{Gao:2020fbl} & $ 155$   & $(109, 610)$ \\
DSE~\cite{Gunkel:2021oya} & $157$   & $(117, 600)$ \\
This work: $2+1$ flavor & $152.3$   & $(102.9, 618.8)$ \\
This work: $2+1+1$ flavor & $151.0$   & $(104.3, 600.1)$ \\
\hline \hline
\end{tabular}
\caption{
The impact of charm quark loop of the CEP location of the chiral phase diagrams together with the results of $N_f=2+1$ in Refs.~\cite{Lu:2023mkn,Fu:2019hdw,Gao:2020qsj,Gao:2020fbl,Gunkel:2021oya}.  Note that here we apply the HTL gluon mass to incorporate the light quark effects in the gluon propagator, which is similar to the approximation applied in Ref.~\cite{Zheng:2024tib}. 
}
\label{tab:ObservablesInPD}
\end{table}


\begin{figure}
    \centering
    \includegraphics[width=1.0\linewidth]{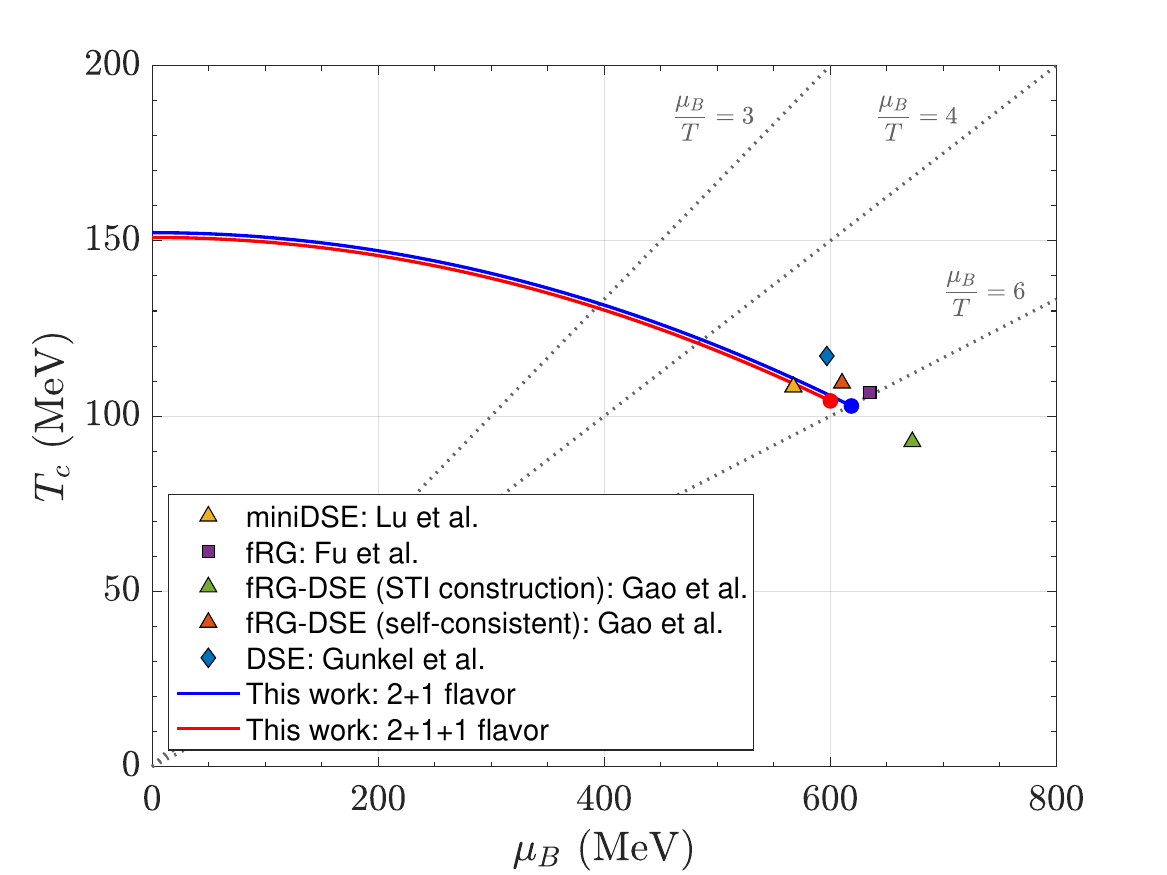}
    \caption{
    QCD chiral phase diagram obtained within the current framework (solid lines) with benchmark results of the critical end point (colored marks).
    The benchmark results come from the Ref.~\cite{Lu:2023mkn} with the miniDSE scheme (yellow triangle), the fRG method~\cite{Fu:2019hdw}, the fRG-DSE scheme with STI construction~\cite{Gao:2020qsj} (green triangle) and with the self-consistent setup~\cite{Gao:2020fbl} (orange triangle), and from the DSE method~\cite{Gunkel:2021oya}.
    The solid lines correspond to the cases of the $2+1$ flavor (blue line) and the $2+1+1$ flavor with the DSE charm loop~\eqref{eq:charmQuarkLoop} (red line).
    }
    \label{fig:PhaseDiagram}
\end{figure}

By scanning the chiral condensate in the temperature and chemical potential plane, we obtain the chiral phase diagram for $N_f=2+1$ and $N_f=2+1+1$. In Fig.~\ref{fig:PhaseDiagram}, we show our results together with the previous studies from functional QCD methods. 
First, we observe that the inclusion of the charm-quark loop has no appreciable impact on the crossover boundary within the present numerical resolution. 
The $N_f=2+1+1$ transition line almost overlaps with the $N_f=2+1$ result throughout the crossover region.
The effect of the charm quark loop on the chiral phase transition is mainly on the location of CEP; that is, the inclusion of the charm quark loop makes the CEP appear at a lower chemical potential. 
The origin of this CEP displacement can be traced to the charm-induced modification of the gluon propagator shown in Fig.~\ref{fig:CharmMassFuncAndGluonDressing}. 
Although the characteristic momentum scale of the gluon dressing function is barely changed, the additional charm-quark loop mildly suppresses the peak of $Z_A(k)$ at intermediate momenta. 
This reduces the effective interaction strength in the momentum region relevant for dynamical chiral symmetry breaking, and consequently, the first-order transition sets in at a slightly lower baryon chemical potential.

Furthermore, we list the values for the  CEP location in Tab.~\ref{tab:ObservablesInPD} in comparison to the previous studies.  After including the charm quark loop, the location of the CEP deviates from the one obtained in the $2+1$ flavor case by a size of around 1.4\% in temperature and around 3.0\% in the baryon chemical potential. 
The values of the CEP location are summarized in Tab.~\ref{tab:ObservablesInPD}. 
Within the same miniDSE setup, the inclusion of the charm-quark loop changes the CEP location from
$(T^{\rm CEP},\mu_B^{\rm CEP})=(102.9,618.8)\,\mathrm{MeV}$
to
$(104.3,600.1)\,\mathrm{MeV}$.
This corresponds to a relative shift of about $1.4\%$ in temperature and about $3.0\%$ in baryon chemical potential. 
Since the two calculations are performed within the same truncation scheme and with the same calibration strategy, this relative displacement provides a direct estimate of the charm-loop effect in the present framework.

The benchmark results from other functional approaches shown in Tab.~\ref{tab:ObservablesInPD} and Fig.~\ref{fig:PhaseDiagram} should be understood as a broader context rather than as a quantitative uncertainty estimate for the charm-loop contribution. 
These approaches differ not only in the treatment of heavy flavors but also in their truncation schemes, input interactions, and dynamical degrees of freedom. 
Nevertheless, their spread illustrates that the charm-induced CEP displacement found here is a modest correction compared with the current model and truncation dependence of CEP predictions.


\section{Summary}
\label{sec:Conclusion}

In this work, we have investigated the impact of charm quark fluctuations on the chiral phase structure of QCD.
Motivated by recent theoretical and experimental developments suggesting that heavy quarks may provide a non-negligible influence on the IR properties of QCD matter, we conducted a comparative analysis between the conventional $2+1$ flavor setup and an extended $2+1+1$ flavor system that includes the dynamical charm quark.

To achieve a controlled and computationally efficient analysis, we employed the recently proposed miniDSE approach, a numerically optimized truncation scheme of the DSEs. 
This framework allows for the consistent treatment of different quark flavors while preserving essential nonperturbative features such as dynamical chiral symmetry breaking. 
Within this setup, we solved the truncated gap equations at finite temperature and baryon chemical potential for both flavor configurations, ensuring all model parameters and renormalization conditions were fixed in the vacuum to allow for direct, self-consistent comparison.

Our results show that the charm-quark loop has only a mild impact on the overall chiral phase boundary. 
In particular, the crossover line obtained in the $2+1+1$-flavor setup almost overlaps with that in the $2+1$-flavor case within the present numerical resolution. 
The main visible effect is instead found in the location of the CEP. 
Within the same miniDSE truncation, the CEP moves from
$(T^{\rm CEP},\mu_B^{\rm CEP})=(102.9,618.8)\,\mathrm{MeV}$
in the $2+1$-flavor case to
$(104.3,600.1)\,\mathrm{MeV}$
after including the charm-quark loop, corresponding to a shift of about $1.4\%$ in temperature and about $3.0\%$ in baryon chemical potential.

The physical origin of this displacement can be traced to the charm-induced modification of the gluon propagator. 
The additional charm-quark loop mildly suppresses the gluon dressing function at intermediate momenta, while leaving the characteristic momentum scale of the gluon propagator almost unchanged. 
This suppression weakens the effective interaction relevant for chiral symmetry breaking, leading to the appearance of the CEP at a slightly lower baryon chemical potential.

We interpret this shift in the phase structure as a manifestation of the charm quark’s indirect yet tangible contribution to the chiral dynamics of QCD. 
We emphasize that the present result should be interpreted as a controlled relative estimate within the miniDSE framework, rather than as a precision determination of the absolute CEP location. 
The direct comparison between the $2+1$- and $2+1+1$-flavor calculations isolates the charm-loop contribution within a fixed truncation and calibration procedure. 
At the same time, the size of the effect is modest when compared with the current spread of CEP predictions from different nonperturbative approaches. 
Therefore, the charm-loop contribution found here should be regarded as a small but systematic correction to the CEP location, while the crossover boundary remains essentially unchanged within the accuracy of the present setup.

In summary, our analysis provides an estimate of the charm-loop contribution to the QCD chiral phase structure within the miniDSE framework. 
The charm quark does not lead to an appreciable deformation of the crossover boundary, but it induces a small and systematic shift of the CEP toward lower baryon chemical potential. 
Future improvements, including more complete quark-gluon vertices and a more self-consistent treatment of the gluon sector, will be required to determine whether such heavy-flavor effects can be resolved beyond the dominant truncation uncertainties.

\begin{acknowledgements}

The work of F.\,G. is supported by the National Science Foundation of China (NSFC) under Grant No. 12305134.
The work of S.\,M. is supported by the NSFC under Grant No. 11747308, 11975108, 12047569, 
and the Seeds Funding of Jilin University (S.M.). 
\end{acknowledgements}





\bibliography{refs}
\end{document}